\documentclass[aps,prd,notitlepage,reprint,twocolumn,showpacs,superscriptaddress,longbibliography,nofootinbib,floatfix]{revtex4-1}

\usepackage{amsmath,amssymb,amsfonts}
\usepackage{graphicx}
\usepackage{verbatim}

\usepackage{hyperref}
\usepackage{slashed}
\usepackage{verbatim}
\newcommand{\be}{\begin{equation}}
\newcommand{\ee}{\end{equation}}
\newcommand{\bi}{\begin{itemize}}
\newcommand{\ei}{\end{itemize}}
\newcommand{\bea}{\begin{eqnarray}}
\newcommand{\eea}{\end{eqnarray}}

\newcommand{\ud}{\mathrm{d}}

\newcommand{\LCm}{{\scriptscriptstyle -}} 
\newcommand{\LCp}{{\scriptscriptstyle +}}

\newcommand{\z}{\boldsymbol{z}}		
\newcommand{\E}{\mathbf{E}}
\newcommand{\B}{\mathbf{B}}

\usepackage[T1]{fontenc} 

\begin{document}

\title{Prospects for studying vacuum polarisation using dipole and synchrotron radiation}

\author{Anton Ilderton}
\email[]{anton.ilderton@chalmers.se}
\affiliation{Department of Physics, Chalmers University, SE-41296 Gothenburg, Sweden}

\author{Mattias Marklund}
\email[]{mattias.marklund@chalmers.se}
\affiliation{Department of Physics, Chalmers University, SE-41296 Gothenburg, Sweden}

\begin{abstract}
The measurement of vacuum polarisation effects, in particular vacuum birefringence, using combined optical and x-ray laser pulses is now actively pursued. Here we briefly examine the feasibility of two alternative setups. The first utilises an alternative target, namely a converging dipole pulse, and the second uses an alternative probe, namely the synchrotron-like emission from highly energetic particles, themselves interacting with a laser pulse. The latter setup has been proposed for experiments at ELI-NP.
\end{abstract}

\maketitle
\section{Introduction}
Light-by-light scattering is a purely quantum effect~\cite{Halpern:1934,Euler:1935zz,Heisenberg:1935qt} which contributes to e.g.~the electron magnetic moment, the Lamb shift and Delbr\"uck scattering. In these cases virtual, or both virtual and real, photons are involved, while light-by-light scattering of only real photons has not yet been observed.
 
While the word `scattering' suggests momentum change, one manifestation of light-by-light effects is the near-{\it forward} scattering of photons with changes to internal degrees of freedom, i.e.~helicity (or polarisation). Consider the collision of two linearly polarised laser pulses, the first a high-intensity optical pulse, the `target', the second a low-intensity X-ray pulse, the `probe'~\cite{Heinzl:2006xc}. Due to the separation in energy scales the probe beam essentially scatters forward, but quantum effects can still cause probe photons to change helicity state. This manifests macroscopically as a slight ellipticity in the probe beam and is hence known as `vacuum birefringence' in analogy to the ellipticity induced in a beam of light passing through a birefringent crystal~\cite{Toll:1952}. Indeed, many phenomena in nonlinear optics have purely photonic analogues, see~\cite{Marklund:2006my,Heinzl:2006xc,DiPiazza:2006pr,KING,Kim:2011fy,Gies:2013yxa,Gies:2014wsa}.

The measurement of vacuum birefringence has been selected as a flagship experiment by the HiBEF consortium at DESY~\cite{HIBEF,HP}, following the proposal in~\cite{Heinzl:2006xc}. For a recent review of the theory behind this topic see~\cite{King:2015tba} and for a detailed review of the current experimental status see~\cite{HP}. 

In this paper we will investigate two alternative, but related, setups which have been suggested for measuring light-by-light effects with real photons. Our goal is simply to obtain a very rough idea of how promising these two alternative schemes are: if they seem promising, the calculations presented here can be refined. The first setup replaces the intense optical laser above with an alternative target, namely an optimally focused ``dipole pulse''~\cite{Ivan}. The second setup retains the intense optical pulse as target, but replaces the above x-ray probe with high-energy photons (gamma rays) emitted as synchrotron radiation from a laser-particle collision. This is the proposed setup for measuring helicity-changing processes at the ELI-NP facility in Romania~\cite{ELI,Nakamiya:2015pde}.

This paper is organised as follows. In Section~\ref{SECT:FLIP} we describe our approach. In Section~\ref{SECT:DIPOL} we consider the alternative target setup. In Section~\ref{SECT:SYNK} we consider the alternative probe and describe the proposed experimental implementation of such a setup at ELI-NP. We conclude in Section~\ref{SECT:CONCS}.

\section{Helicity-flip in background fields}\label{SECT:FLIP}
Recall the standard optics result for the polarisation ellipticity $\delta$ induced in a beam of light, frequency $\omega'$, passing through a birefringent medium of length $d$, refractive indices $\{n_\LCp,n_\LCm\}$:
\be\label{est1}
	\delta = \frac{1}{2}(n_+ - n_-)\omega' d \;.
\ee
The quantum vacuum exposed to a strong field effectively develops `vacuum refractive indices' which arise through the nonlinearity of the Euler-Heisenberg action~\cite{Euler:1935zz,Heisenberg:1935qt}, and can be calculated using the photon polarisation tensor. In the limit that the strong field is a constant, homogeneous crossed field of strength $E$, a counter-propagating probe sees the indices~\cite{Toll:1952,Narozhnyi:1968}
\be \label{N.PM1}
  n_\pm = 1 +  \frac{\alpha}{45 \pi} (11 \pm 3) \frac{E^2}{E_S^2} \;,
\ee
where $E_S = m^2/e \simeq 10^{18}$ V/m is the Sauter-Schwinger field. Inserting (\ref{N.PM1}) into (\ref{est1}) we obtain the ellipticity induced in the probe as
\be\label{est2}
 	\delta \to \frac{\alpha}{15\pi}\frac{E^2}{E_S^2}\omega' d \;.
\ee
This macroscopic beam ellipticity induced by quantum effects is `vacuum birefringence';  the microscopic physics underlying it is as follows.

Consider a probe photon, momentum $l_\mu \equiv \omega' \hat{l}_\mu$, frequency $\omega'$, and helicity state described by $\epsilon_\mu$. The photon passes through a strong background field $F_{\mu\nu}$ with typical frequency scale much smaller than $\omega'$ (as would be the case for an X-ray probe of an optical laser). In this case scattering becomes essentially forward. The probability $\mathbb{P}$ that the probe photon flips to its opposite helicity state $\epsilon'_\mu$ may be written $\mathbb{P}_\text{flip} = |\mathbb{T}|^2$, where the amplitude $\mathbb T$ can be approximated by a line-integral over the classical (i.e.~straight line) trajectory of the photon~\cite{Dinu:2014tsa}, here parameterised with time $t$:
\be\label{T1}
	\mathbb{T} =\frac{\alpha}{30}\frac{\omega'}{E_S^2} \int\!\ud t\; \big( \bar{\epsilon}'_\mu F^{\mu\nu}\hat{l}_\nu\big) \big( {\epsilon}_\sigma F^{\sigma\rho}\hat{l}_\rho\big) \;.
\ee
$F_{\mu\nu}$ is evaluated on the photon trajectory.  The probability is maximised when the field and probe polarisations can be chosen to lie at a relative angle of $45^\circ$. $\mathbb{T}$ then reduces to 
\be\label{T2}
	\mathbb{T} = \frac{\alpha}{60}\frac{\omega'}{E_S^2} \int\! \ud t\  \hat{l}_\mu T^{\mu\nu} \hat{l}_\nu \;,
\ee
where $T_{\mu\nu}$ is the background field energy-momentum tensor; thus we can interpret (\ref{T2}) as simply being proportional to an integrated energy density (an intensity) seen by the probe as it passes through the target~\cite{Dinu:2014tsa}. That this is an integrated, rather than peak, variable will be important below.

As detailed in~\cite{Dinu:2013gaa}, the flip probability $\mathbb{P}$ is directly related to the ellipticity~$\delta^2$: hence (\ref{T1}) is most easily interpreted as a quantum field theory generalisation of the classical result (\ref{est1}), which goes beyond (\ref{est2}) as it  allows us to consider arbitrary field strengths and shapes (on the usual provisos that the field strength is not of Schwinger scale and the invariant, c.o.m., frequency scales do not exceed the electron mass).

There are several effects which we do not include in this first investigation. No depletion of the background field is accounted for, nor do we account for probe scattering~\cite{Lundstrom:2005za,King:2012aw}, for an investigation of which in vacuum birefringence see~\cite{Karbstein:2015xra}. We also restrict our attention to single photon probes; beam-like probes can be accounted for using a straightforward extension of the formalism used here~\cite{Dinu:2014tsa,Torgrimsson:2014sra}. In summary, our current approximation gives a good estimate for the on-axis briefringence signal. (Note though that in experiments with either laser or magnetic fields scattered photon signals may be easier to detect than on-axis signals, due to lower backgrounds~\cite{Gies:2013yxa,Karbstein:2015qwa,Karbstein:2015xra}.)
%
\subsection{Conventions and notation}
%
The helicity-state vectors for a photon of momentum $l_\mu$ are
\be
	\epsilon_\pm^\mu = \frac{1}{\sqrt{2}} \big( \epsilon_1^\mu \pm i \epsilon_2^\mu \big) \;,
\ee
where, using Coulomb gauge,
\be
\begin{split}
	l^\mu &= \omega' (1, \sin\theta \cos\phi , \sin\theta \sin\phi , \cos\phi) \;, \\
	\epsilon_1^\mu &= (0,\cos\theta \cos\phi, \cos\theta \sin\phi, -\sin\phi) \;, \\
	\epsilon_2^\mu &=(0,-\sin\phi , \cos\phi, 0) \;.
\end{split}
\ee
%
%
%
\section{Dipole pulse targets}\label{SECT:DIPOL}
%
Dipole pulses are exact, singularity-free, optimally focussed, finite-energy solutions of Maxwell's equations in vacuum~\cite{Ivan}. In an ``$e$-dipole'' pulse the electric field dominates over the magnetic field in the focus, and provides optimal conditions for pair production via the nonperturbative Sauter-Schwinger mechanism~\cite{Gonoskov:2013ada,Gonoskov:2013aoa}. In an ``$h$-dipole'' pulse, the magnetic field dominates and one might ask whether the optimal focussing amplifies the helicity-flip probability. To investigate this we consider replacing the intense beam in the vacuum birefringence experiments described above with an \textit{h}-dipole pulse.

The fields of an \textit{h}-dipole pulse are written in terms of a function $\mathbf Z$ defined by
\be\label{Z-DEFN}
	\mathbf{Z} = \hat{\z}\,\frac{d}{|{\bf x}|}  \big[{g(t+|{\bf x}|)}-{g(t-|{\bf x}|)}\big] \;,
\ee
where the `driving function' $g$ will be specified shortly and the `virtual dipole moment' $d$ is a constant. The fields are
\be\label{F-DEFN}
	\B = -\nabla\times\nabla\times {\mathbf{Z}}, \quad \E = \nabla\times \dot{\mathbf{Z}} \;,
\ee
and in the focus are equal to
\be\label{B0}
	\B(0,t)=\hat{\z}\frac{4d}{3}\dddot{g}(t) \;,\quad \E(0,t)={0} \;.
\ee
For the driving function we choose a Gaussian,
\be\label{QUASI-GAUSS}
	g(t) = e^{-\Delta\omega^2t^2/4}\sin(\omega t) \;,
\ee
in which $\omega$ is the central frequency and $\Delta\omega$ is a frequency spread. In the focus we have, from (\ref{B0}), the same frequency spread as in $g$. The intensity distribution of the dipole pulse has the form~\cite{Ivan}
\be\label{int}
	I=I_0\sin^2\theta \;,§
\ee
where $\theta$ is the angle made with the $z$-axis.

\begin{figure*}
\includegraphics[width=0.9\columnwidth]{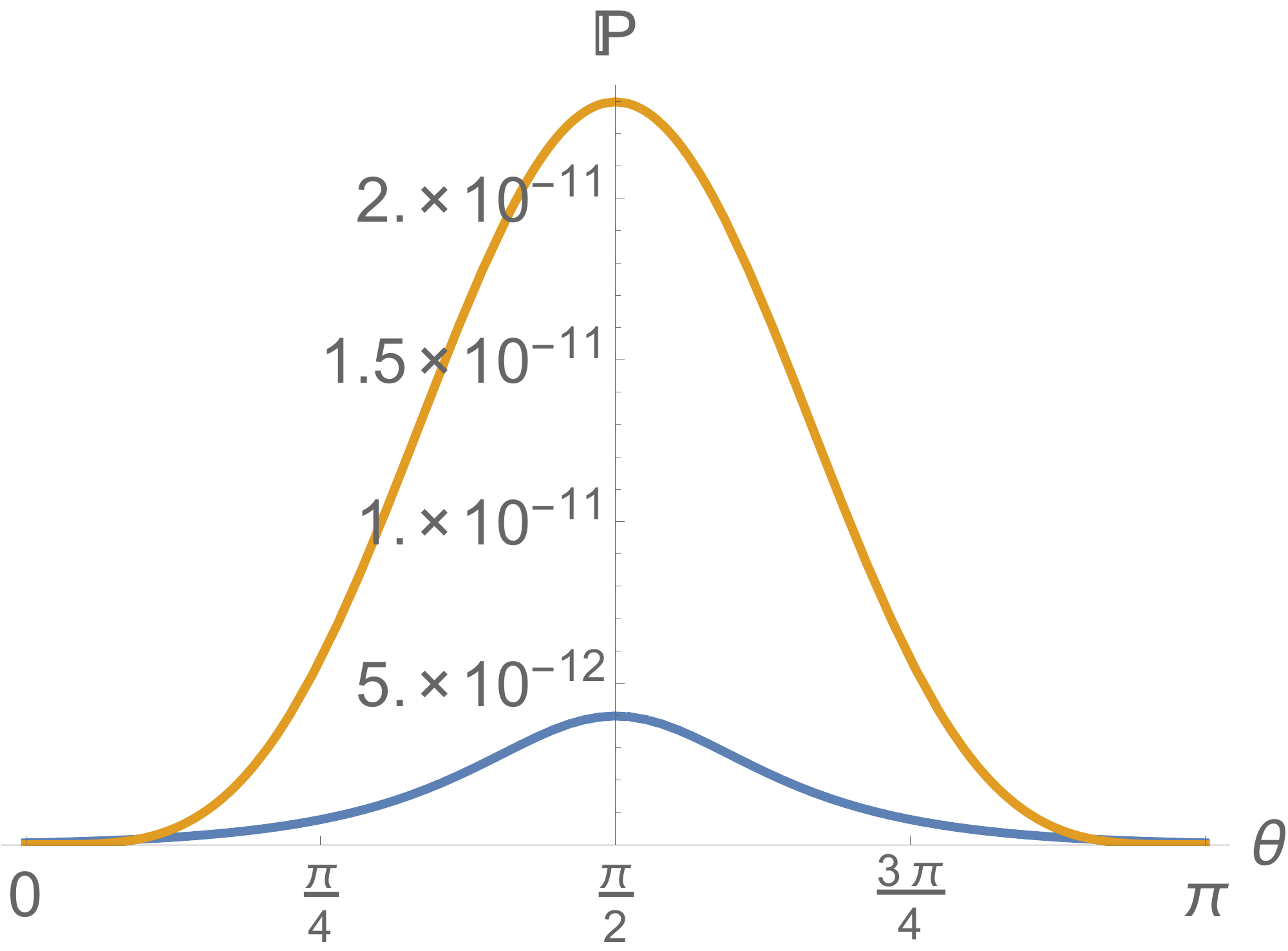}
\hspace{10pt}
\raisebox{5pt}{\includegraphics[width=0.9\columnwidth]{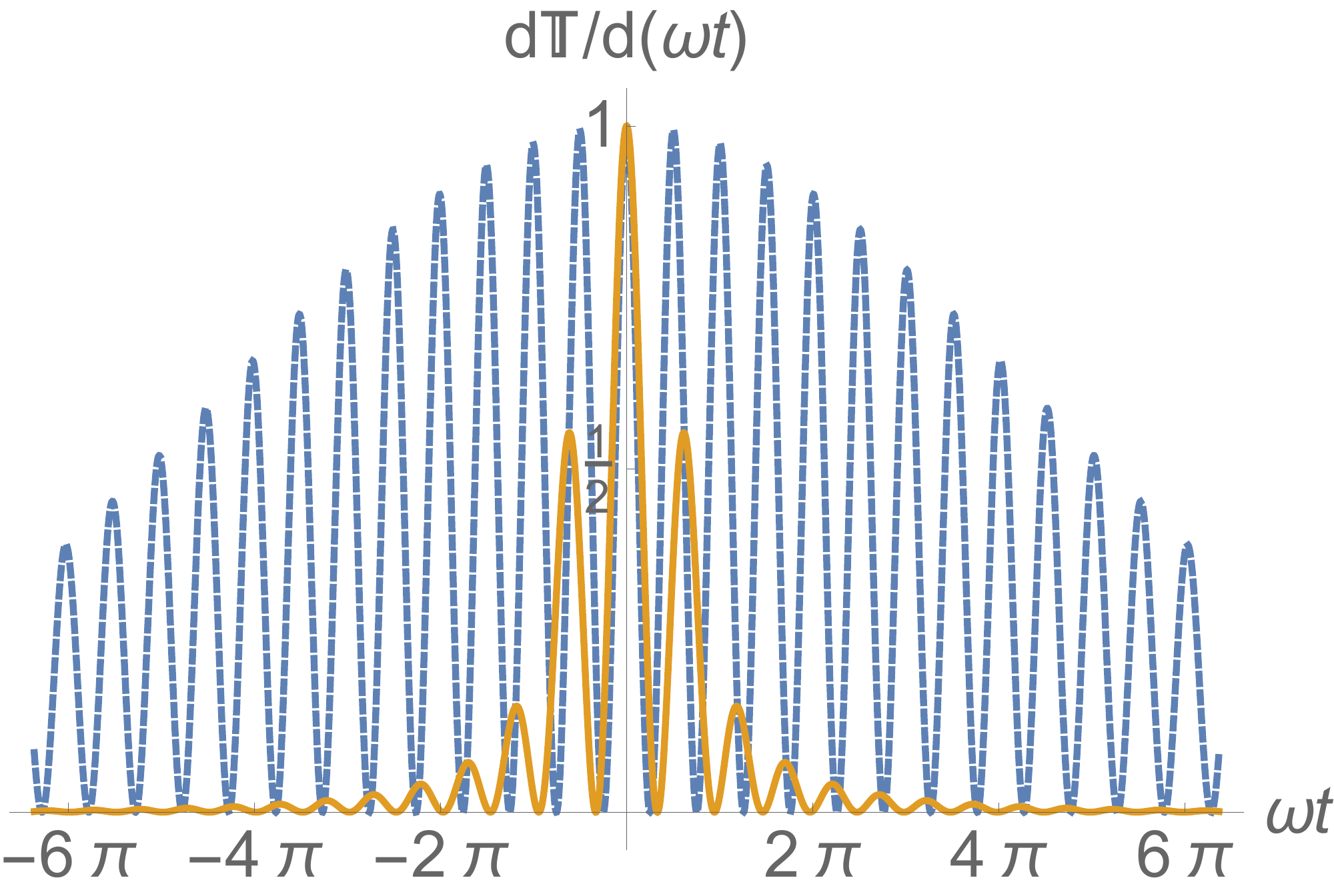}}
\caption{\label{FIG:FAS} \textit{Left:} Helicity flip probability $\mathbb{P}$ in the dipole (yellow) and Gaussian (blue, with $\theta$ shifted by $\pi/2$ to be able to plot on the same scale). \textit{Right:} The integrands of the \textit{normalised} probability amplitudes $\mathbb{T}$; in the Gaussian the focal spot is larger, and it can be seen that $\mathbb{T}$ indeed receives contributions from a larger time interval.}
\end{figure*}

We will compare the flip probability in a dipole pulse with that in a Gaussian (paraxial) beam of the same input energy, using the expected parameters for the PW laser at DESY in conjunction with birefringence experiments~\cite{HP}.  We take a total energy of $30$~J, wavelength $\lambda=800~$nm, and a bandwidth of $\Delta\omega \simeq 0.035\omega$ (corresponding to a FWHM pulse duration $28\,$fs). For the Gaussian we also need to choose a focal spot radius, which we take to be $w_0=1.75\mu m$, again following~\cite{HP}.

For these parameters the peak fields in the foci of the dipole and Gaussian beams become
\be\label{dens}
	\frac{1}{2}({\bf E}^2+ {\bf B}^2) \simeq 
	\begin{cases}
		3 \times 10^{-6}E_S^2 & \text{Dipole}\;, \\
		9 \times 10^{-8}E_S^2 & \text{Gaussian}\;,
	\end{cases}
\ee
differing by over an order of magnitude: we may therefore expect a significantly stronger birefringence signal in the dipole pulse than in the Gaussian beam. However other factors also play a roll, e.g.~polarisation alignment between probe and target (the dipole pulse is radially polarised). Turning to the flip probability (\ref{T1}) or (\ref{T2}) we also need a probe photon trajectory. We consider the best possible scenario where the probe passes through the field focus (i.e.~zero impact parameter) at the instant of peak field strength (i.e.~no timing miss). The resulting probabilities (naturally) follow the intensity profiles of the background fields. That in the dipole, for example, follows (\ref{int}) and has the form
\be
	|\mathbb{T}_\text{flip}|^2 = C_0 \sin^4\theta \;,
\ee
where the incoming probe makes an angle $\theta$ to the $z$-axis and $C_0$ depends on the dipole pulse parameters and probe frequency, but not on $\theta$ or polarisation angles.

Further explicit expressions are unrevealing -- instead we plot in Fig.~\ref{FIG:FAS} the flip probability in the dipole pulse and in the Gaussian beam. The probability in the dipole pulse exceeds that in the Gaussian by a factor of around~$5.8$, which is less than might be expected from (\ref{dens}). The reason for this is seen by recalling that it is an \textit{integrated} parameter to which birefringence is sensitive, and that while the focal field strength in a dipole pulse is higher than in a Gaussian, the spot size is smaller. We can confirm this by estimating the effective transverse extent of the focus in our dipole pulse (transverse since the best-case scenario is for probe angle $\theta=\pi/2$). Following~\cite{Ivan} we define the effective extent as the distance from the focal point at which the energy density drops to half its peak value. For our parameters we find a sub-wavelength extent $\simeq 0.4\lambda$. In Fig.~\ref{FIG:FAS}, right panel, we plot the (normalised) integrands of $\mathbb{T}$ in the dipole and Gaussian beams. We clearly see that the probability amplitude receives contributions from a much larger phase range in a Gaussian beam than it does in a dipole pulse; for the dipole case the width of the central peak is roughly $0.4\lambda$, consistent with expectations.

\section{Synchrotron emission as a probe}\label{SECT:SYNK}
\begin{figure}
\centering\includegraphics[width=\columnwidth]{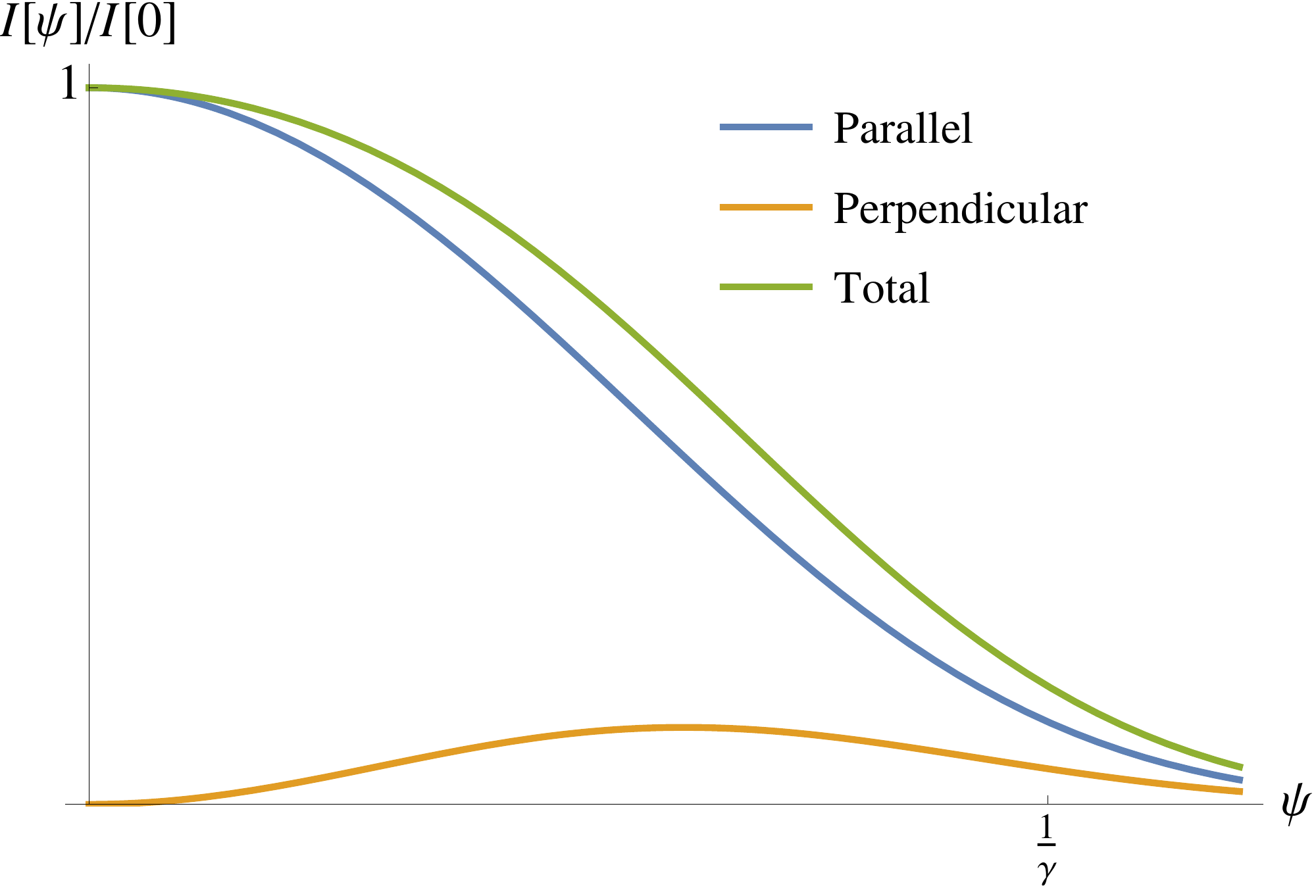}
\includegraphics[width=\columnwidth]{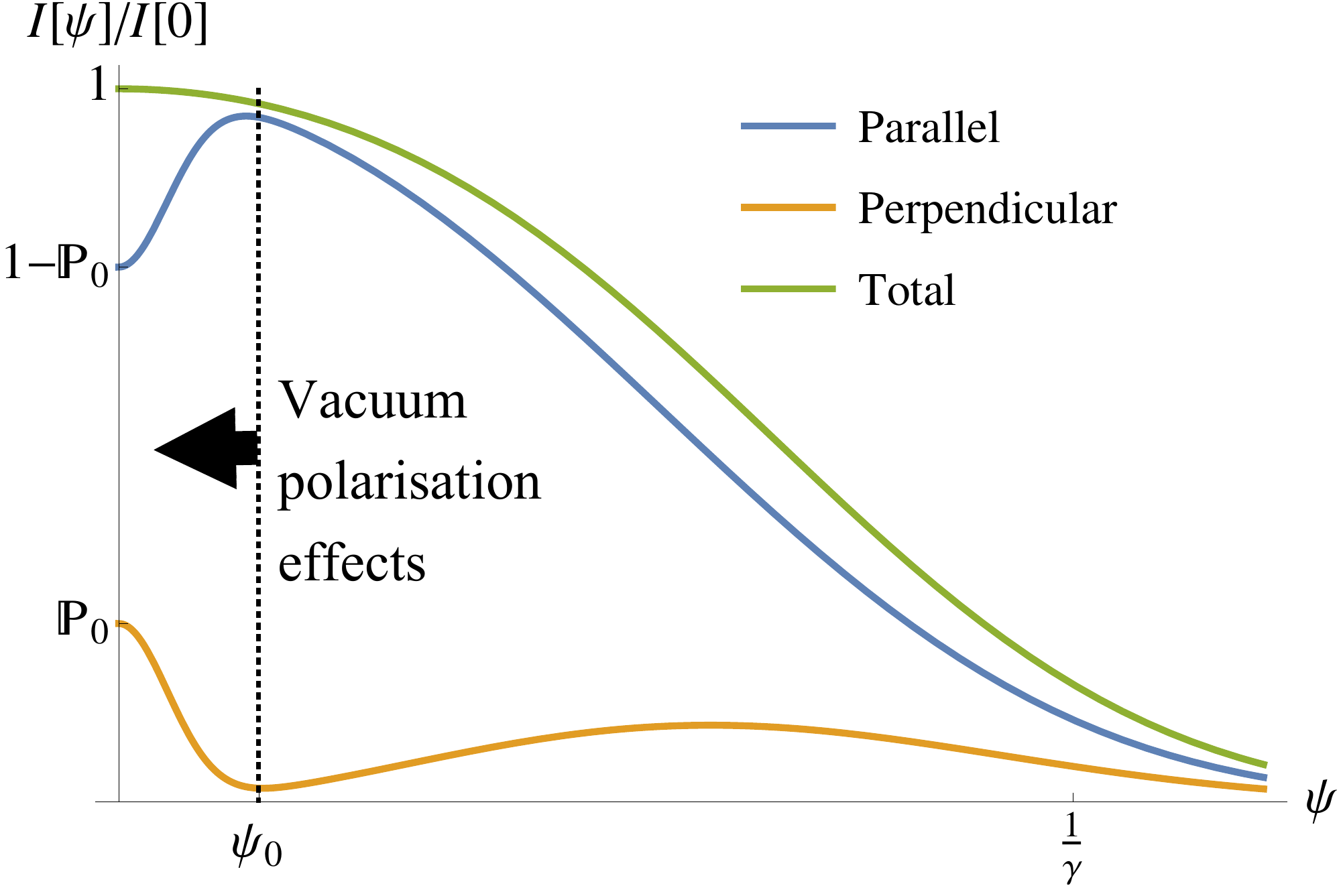}
\caption{\label{FIG:SYNK} {\it Top}: Standard synchrotron emission spectrum as a function of opening angle $\psi$ at, to illustrate, $\omega'=1.3\omega_c$. The radiation is emitted in a narrow cone of opening angle $\psi\sim1/\gamma$. {\it Bottom}: The same spectrum illustrating the effects of vacuum polarisation, which are confined to narrower angles $\psi\lesssim \psi_0$ defined by the interaction geometry. Photon helicity-flip mixes the parallel and perpendicularly-polarised components of the synchrotron spectrum.}
\end{figure}

%
Above we discussed an ``alternative target'' for measuring vacuum birefringence. We now turn to an ``alternative probe'', namely synchrotron emission. We begin by recalling some standard results~\cite{Duke:2000wc}. The spectral density of synchrotron emission from a particle with gamma-factor $\gamma$ moving in planar circular motion, radius $R$, is
\be\label{II}
	I = I_0 \gamma^2 \bigg(\frac{\omega}{\omega_c}\bigg)^2
	(1+\gamma ^2 \psi^2)^2 
	\bigg( K^2_\frac{2}{3}(\xi) + \frac{\gamma^2 \psi^2}{1+\gamma^2 \psi^2} K^2_\frac{1}{3}(\xi) \bigg) \;,
\ee
where the critical frequency is $\omega_c =3\gamma^3/2R $, $\psi$ is the angle of elevation out of the plane of motion, $\xi = (\omega/\omega_c)(1+\gamma^2\psi^2)^{3/2}$ and $I_0$ is an overall normalisation which is not important here. The two terms in the large brackets of (\ref{II}) represent, respectively, the intensities radiated parallel and perpendicular to the plane of motion, which we write as $I_\parallel$ and $I_\perp$. Synchrotron radiation is highly plane-polarised, as illustrated in Fig.~\ref{FIG:SYNK}. The small-angle part of the spectrum is therefore a potential source of highly polarised photons for use in birefringence experiments: if these photons interact with an intense optical pulse, helicity-flip will mix the plane- and perpendicular-polarised parts of the emitted radiation, `deforming' the synchrotron spectrum. (Of course we need a high energy synchrotron spectrum to obtain an appreciable flip probability, see below.) 
\begin{figure}[b]
	\centering\includegraphics[width=0.9\columnwidth]{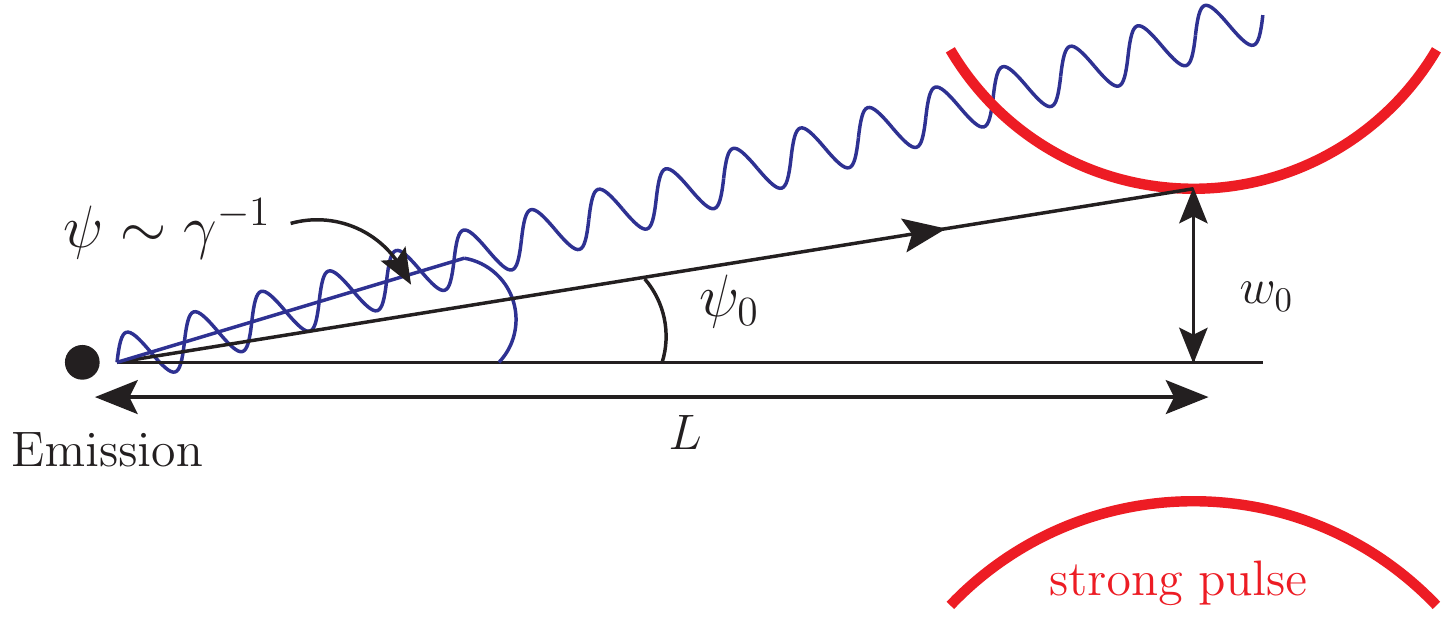}
	\caption{\label{FIG:TRI} Sketch of experimental geometry, showing the distance between photon emission and interaction points. Emission is near-forward, $\psi < 1/\gamma$, while the effective emission range of photons which can interact with the high-intensity pulse is limited to $\psi <\psi_0:=\tan^{-1}L/w_0$.}
\end{figure}

The portion of photons which will interact with the focal spot of the intense pulse is limited by the geometry of the experiment. Assume that the distance between the emission point of the high-energy photons and the focal point of the high-intensity pulse is $L$, and that the pulse's focal width is $w_0$ -- see Fig.~\ref{FIG:TRI}. Clearly only photons emitted in a very narrow angle $\psi<\psi_0 := \tan^{-1} w_0/L$ will see the laser focal spot and be likely to change helicity state\footnote{We consider only photons which arrive at the focal spot at the instance of peak field strength -- for the impact of timing jitter see~\cite{Dinu:2014tsa,HP}.}. This will be, as we verify below, much smaller than the typical opening angle ($1/\gamma$) of the synchrotron spectrum, so vacuum polarisation effects will only be observable for photons emitted almost within the plane of motion of the electrons.

Given a setup as in Fig.~\ref{FIG:TRI} we can use (\ref{T1}) to calculate the flip probability $\mathbb{P}$.  Let $\mathbb{P}_0$ be the `best case' flip probability for photons arriving at the focal spot at the instant of peak field strength and with polarisation at $45^\circ$ to that of the intense optical pulse. Now consider deviations from the ideal case: the dependence of the probability on emission angle $\psi$ is easily guessed as being Gaussian, since the probability is expected to follow the intensity distribution squared. Indeed
\be\label{typ}
	\mathbb{P}(\omega',\psi) = \mathbb{P}_0(\omega') e^{ -4\psi^2/\psi_0^2 } \;,
\ee
gives a very good approximation of the flip probability: additional dependencies on geometric or polarisation angles due to perturbing away from the ideal case are effectively washed out by the very rapid falloff of the probability with $\psi$. With this we can write down a simple model of the synchrotron emission spectrum following interaction with an intense pulse. Denoting the outgoing distribution with a prime, we write
\be\begin{split}\label{UT}
	I'_\parallel &= (1- \mathbb{P}) I_\parallel + \mathbb{P} I_\perp \;, \\
	I'_\perp &= (1- \mathbb{P}) I_\perp + \mathbb{P} I_\parallel \;,
\end{split}
\ee
(implying no photons are lost: $I' := I'_\parallel + I'_\perp = I$). Vacuum polarisation then has a significant impact on the spectrum only for $\psi < \psi_0$, as is illustrated in the bottom panel of Fig.~\ref{FIG:SYNK}. With this model in hand we turn to quantitative estimates for a proposed experiment at ELI-NP~\cite{Nakamiya:2015pde}.

\subsection{Experiments at ELI-NP}
The ELI proposal begins with the collision of highly relativistic electrons, $\gamma\gg 1$, with a linearly polarised laser pulse. The electrons undergo Compton back-scattering and emit instantaneously in a synchrotron spectrum~\cite{Jackson:1998nia}.  The polarisation direction of the emitted radiation is set by the plane of motion of the electrons, which is in turn set by the laser polarisation direction. The angle $\psi$, above, is the angle out of this plane.

The produced high-energy photons are then used as the probe of a (second) laser pulse of very high intensity. The combination of high intensity and high-energy increases the probability of helicity flip as the probe photons pass through the optical laser, see~(\ref{T1})-(\ref{T2}). After this interaction the polarisation of the high-energy photons is measured using a pair-polarimeter~\cite{Nakamiya:2015pde}, see also below.

We assume generation of $2$~GeV electrons which are collided with a laser pulse of by modern standards moderate intensity $I\sim 10^{20}$~W/cm$^2$. This generates radiation with a critical frequency of $\omega_c = (3/2)m\gamma^2(E/E_S) = 0.24$~GeV (using $I=E^2/2$) which is to interact with an intense optical pulse. 

We assume a distance of $L=20$~cm between the emission point of the radiation and the interaction point with the intense pulse. Based on expected ELI parameters we take a focal radius $w_0=2.5\,\mu$m. This gives the effective emission angle as $\psi_0 = \tan^{-1} 2.5\mu \mathrm{m}/ 20 \mathrm{cm} \simeq 10^{-5}$, which is as suggested above much less than $1/\gamma\sim 3\times 10^{-4}$. In order to write down the flip probability, we need the `best case' result as described above (\ref{typ}). For a focussed Gaussian beam this has been found in~\cite{Dinu:2014tsa} to be
\be\label{P0}
	\mathbb{P}_0(\omega') = \bigg( \frac{\alpha}{15}
	\frac{1}{E_S^2}
	\frac{\mathcal{E}\omega'}{\pi^2 w_0^2}
	\bigg)^2 \;.
\ee
where $\mathcal{E}$ is the energy of the laser. Based on expected ELI parameters we take $\mathcal{E}=200\,\mathrm{J}$, which gives
\be
	\mathbb{P}_0(\omega') \simeq 0.27\ \bigg(\frac{\omega'}{\text{GeV}}\bigg)^2 \;.
\ee
This is significantly higher than for an optical--X-ray setup simply due to the higher probe energy (and the expected higher energies and intensities available at ELI). The flip probability as a function of $\psi$ is then given by (\ref{typ}). 

\begin{figure}
	\centering\includegraphics[width=0.9\columnwidth]{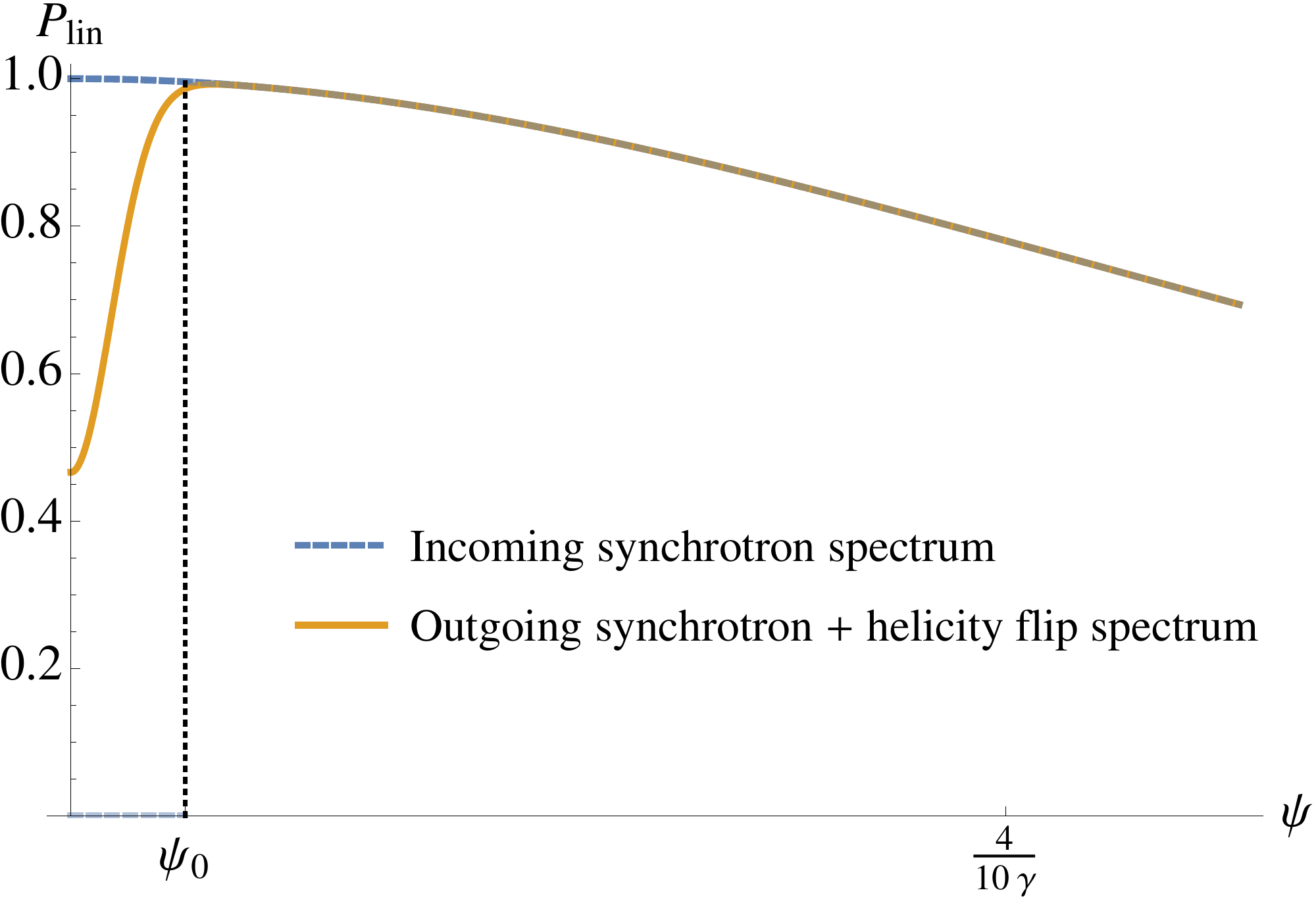}
	\caption{\label{FIG:PLIN} The degree of linear polarisation in the synchrotron spectrum (blue/dashed) and in the spectrum after passing through an intense field (yellow/solid) in which vacuum polarisation effects cause changes in photon polarisation. Plotted for $1$~GeV probe photon energy and other parameters as in the text, for the proposed setup at ELI-NP.}
\end{figure}

In the proposed experiment, the emitted radiation will be screened in order to ensure a high polarisation purity; only that part of the spectrum emitted at $\psi$ less than some small fraction, say 40\%, of $1/\gamma$ will be allowed to propagate toward the high intensity pulse (the target). A detector will be arranged to screen out (to some high degree) plane-polarised emission. The signal to be measured is then the increase in perpendicularly-polarised photons incident on the detector due to vacuum polarisation effects. We can calculate the total energy deposited on the detector due to perpendicularly polarised photons from\footnote{We integrate only up the electron energy, as quantum effects will cut off the spectrum there: this and other such refinements should be included in future calculations.}~\cite{Jackson:1998nia}
\be
	\text{En}_\perp' = \int\limits_0^{0.4/\gamma}\!\ud\psi \! \int\limits_0^\text{2 GeV}\!\ud \omega'\;  I_\perp'(\omega',\psi) \simeq 1.16\,  \text{En}_\perp \;,
\ee
which gives an increase of 16\%.  A convenient measure of polarisation purity (which is also related to the polarimetry required to measure the photon polarisation in this setup~\cite{Nakamiya:2015pde}) is the ``degree of linear polarisation'', which for the synchrotron spectrum before and after interaction with the intense laser pulse is defined by
\be
	P_\text{lin} := \frac{I_\parallel - I_\perp}{I_\parallel + I_\perp}\;, \qquad P^\prime_\text{lin} = \frac{I'_\parallel - I'_\perp}{I'_\parallel + I'_\perp} = \big( 1 - 2 \mathbb{P}\big) P_\text{lin} \;.
\ee
Vacuum polarisation effects will (for $\mathbb{P}<0.5$) reduce the degree of linear polarisation: this is illustrated for the parameters considered here in Fig.~\ref{FIG:PLIN}.

%
\section{Discussion and conclusions}\label{SECT:CONCS}
%
We have considered two proposals for measuring vacuum polarisation effects in strong laser fields. In the first, we used a dipole pulse as the `target'. The optimal focussing of dipole pulses yields high focal field strengths, which makes them ideal for studying pair creation~\cite{Gonoskov:2013ada} and intense-field dynamics~\cite{Gonoskov:2013aoa}. However the dipole pulse has a small (sub-wavelength) focal spot size, which can be disadvantageous for vacuum birefringence as the relevant observable there is sensitive to, essentially, the product of field strength and spot size.

The second method we have considered is the use of laser-particle collisions to generate high energy gamma rays, which are in turn used as the probe of an intense optical pulse. We have provided a very simple `proof of principle' calculation and seen that the high energy of the probe photons gives a large (ideal case) helicity flip probability. The experimental realisation of the relevant setup though will be challenging. Measurement of the signal requires pair polarimetry on the probe gamma rays; this is discussed in ~\cite{Nakamiya:2015pde}. The use of a laser-particle collision to generate the probe in close proximity to the target suggests a `messy' experimental environment. Only photons generated in a small volume of space, at the right time, will interact with the focal spot of the intense pulse and have an appreciable chance of changing helicity state; however the actual generation point can be anywhere in the volume of the laser-particle collision. This suggests that shot-shot fluctuations in signal and background may be large.

Refinements of the calculation presented here could begin with numerical simulations of the initial laser-particle interaction in order to better understand the spectrum of the generated probe photons~\cite{King2016}. Here PIC methods would be useful, for a review of which see~\cite{Gonoskov:2014mda}. (Polarisation effects would of course need to be included.) Once the spectrum is understood the impact of effects such as timing and pointing jitter can be included, and then a comprehensive picture of the background and signal sizes can be developed.

{\it The authors are supported by the Olle Engkvist Foundation, grant 2014/744 (A.I.) and the Swedish Research Council, grants 2012-5644 and 2013-4248 (M.M). A.I.~thanks the coordinators of ELI-NP for the opportunity to join the facility's Technical Design Report.}

\end{document}